\documentclass{iau}
\usepackage{graphicx}

\title[New version of the Balona light curve modelling technique]
{The Baade-Becker-Wesselink technique and the fundamental
astrophysical parameters of Cepheids}

\author[A. S. Rastorguev et al.]
{Alexey S. Rastorguev$^1$, Andrey K. Dambis$^1$, Marina V.
Zabolotskikh$^1$, Leonid N. Berdnikov$^1$, and Natalia A.
Gorynya$^1$}

\affiliation{$^1$Lomonosov Moscow State University, \\ Sternberg
Astronomical Institute, \\ 13 Universitetskii prospect, Moscow, 119992 Russia \\
email: {\tt rastor@sai.msu.ru}}

\pubyear{2013}
\volume{289}  
\pagerange{1--8}
\jname{Advancing the physics of cosmic distances}
\editors{R. de Grijs and G. Bono, eds.}
\begin{document}

\maketitle

\begin{abstract}
The BBW method remains one of most demanded tool to derive full
set of Cepheid astrophysical parameters. Surface brightness
version of the BBW technique was preferentially used during last
decades to calculate Cepheid radii and to improve PLC relations.
Its implementation requires a priory knowledge of Cepheid
reddening value. We propose a new version of the
Baade--Becker--Wesselink technique, which allows one to
independently determine the colour excess and the intrinsic colour
of a radially pulsating star, in addition to its radius,
luminosity, and distance. It is considered to be a generalization
of the Balona light curve modelling approach. The method also
allows calibration of the function $F(CI_0) = BC(CI_0) + 10 \times
log (T_{eff} (CI_0))$ for the class of pulsating stars considered.
We apply this technique to a number of classical Cepheids with
very accurate light and radial-velocity curves. The new technique
can also be applied to other pulsating variables, e.g. RR~Lyraes.
We also discuss the possible dependence of the projection factor
on the pulsation phase.

\end{abstract}

\firstsection 

\section{Introduction}

Classical Cepheids are the key standard candles. They are used to
set the zero point of the extragalactic distance scale
\cite[(Freedman~et~al~2001)]{F01} and also serve as important
tracers of young populations \cite[(Binney and
Merrifield~1998)]{BM}. They owe their popularity to their high
luminosities and photometric variability (which make them easy to
identify and observe even at large distances) and the fact that
the luminosities, intrinsic colours, and ages of these stars are
closely related to such an easy-to-determine quantity as the
period of variability.

It would be best to calibrate the Cepheid period-luminosity (PL),
period-colour (PC), and period-luminosity-colour relations using
distances based on trigonometric parallaxes. However, the most
precisely measured parallaxes of even the nearest Cepeheids remain
insufficiently accurate and, more importantly, they may be fraught
with so far uncovered systematic errors. Here the
Baade--Becker--Wesselink method \cite[(BBW; Baade~1926,
Becker~1940, Wesselink~1946)]{Baa26, Beck40, Wes46} comes in
handy, because it allows the Cepheid distances (along with the
physical parameters of these stars) to be inferred, thereby
providing an independent check of results based on geometric
methods (e.g., trigonometric and statistical parallaxes). The
surface brightness technique was used most frequently and
effectively during the past few decades. It is based on the
relation between so-called limb-darkened surface brightness
parameter and the normal colours of Cepheids \cite[(Barnes and
Evans~1976)]{BE76}. Moreover, it critically depends on the adopted
reddening value.

Cepheid reddening values can be estimated from medium- and
broad-band photometric observations (including multicolour PL
relations) and from spectroscopic data \cite[(Dean, Warren and
Cousins~1978, Fernie~1987, Fernie~1990,~1994, Fernie~et~al.~1995,
Berdnikov, Vozyakova and Dambis~1996, Berdnikov, Dambis and
Vozyakova~2000, Andrievsky~et~al.~2002a,~b, Kovtyukh~et~al.~2008,
Kim, Moon and Yushchenko~2011)]{DWC78, Fer87, Fer90, Fer94,
Feretal95, Berd96, Berd00, Andr02a, Andr02b, Kov08, KMY11}. All
these methods use proper calibrations and relationships between
key stellar parameters. However, there exist large (up to $0.2
mag$) scatter in the estimates of the colour excess for individual
Cepheids, and the structure of the Cepheid instability strip is
still vague. In the review on the Hubble constant and the Cepheid
distance scale \cite[Madore and Freedman~(1991)]{MF91} noted that
``...any attempt to disentangle the effects of differential
reddening and true color deviations within the instability strip
must rely first on a precise and thoroughly independent
determination of the intrinsic structure of the
period-luminosity-color relation.'' And next: ``...independent
reddenings and distances to individual calibrator Cepheids must be
available.''

It should also be noted that reliable values of the colour excess
and of the total-to-selective extinction ratio, say, $A_V /
E(V-I)$, are extremely important when we use Wesenheit function to
derive Cepheid luminosities from precise trigonometric parallaxes
\cite[(Groenewegen and Oudmaijer~2000, Sandage~et~al.~2006, van
Leewen~et~al.~2007)]{GO00, Sand06, vL07}. To convert Wesenheit
index $W_{VI}$ to the absolute magnitude, $M_V$, the intrinsic
colour of the Cepheid, $(V-I)_0$, and the proper value of $A_V /
E(V-I)$ are needed. We suppose that self-consistent and
independent reddening estimates for individual Cepheids can reduce
underestimated systematical errors induced by large variations of
the absorption law \cite[(Fitzpatrick and Massa~2007)]{FM07} and
can even result in $A_\lambda$ law in optics and NIR. Therefore,
the search for independent estimates of Cepheid reddening values
is still actual, and this is our primary aim.

Both BBW techniques -- surface brightness (radius-variation
modelling) and maximum likelihood (light-curve modelling) -- are
based on the same astrophysical background but make use somewhat
different calibrations (limb-darkened surface brightness
parameter, bolometric correction -- effective temperature pair) on
the normal colours. Here we propose a generalization of the
\cite[Balona~(1977)]{B77} light-curve modelling technique, which
allows one to independently determine not only the star's distance
and physical parameters, but also the amount of interstellar
reddening, and even calibrate the dependence of a linear
combination of the bolometric correction and effective temperature
on intrinsic colour \cite[(Rastorguev and Dambis~2011)]{RD11}.

\section{Theoretical background}

We now briefly outline the method. First, the bolometric
luminosity of a star at any time instant is given by the following
relation, which immediately follows from the Stefan--Boltzmann
law:
\begin{equation}
L/L_{\odot} = (R/R_{\odot})^2 \times (T/T_{\odot})^4.
\end{equation}
Here $L$, $R$, and $T$ are the star's current bolometric
luminosity, radius, and effective temperature, respectively, and
the $\odot$ subscript denotes the corresponding solar values.
Given that the bolometric absolute magnitude, $M_{bol}$, is
related to bolometric luminosity as
$$
M_{bol} = M_{{bol}\odot} - 2.5 \times log (L/L_{\odot}),
$$
we can simply derive from Eq. (2.1):
\begin{equation}
M_{bol} - M_{{bol}\odot} = -5 \times log (R/R_{\odot})- 10 \times
log (T/T_{\odot})
\end{equation}
Now, $M_{bol}$ can be written in terms of the absolute magnitude
$M$ in some photometric band and the corresponding bolometric
correction, $BC$, i.e.
$$
M_{bol} = M + BC,
$$
and the absolute magnitude $M$ can be written as:
$$
M = m - A - 5 \times log (d/10~pc).
$$
Here $m$, $A$, and $d$ are the star's apparent magnitude and
interstellar extinction in the corresponding photometric band,
respectively, and $d$ is the heliocentric distance of the star in
pc. We can therefore rewrite Eq. (2.2) as follows:
$$
m=A+5\times log (d/10~pc)+M_{{bol}\odot}+10\times log(T_{\odot})
$$
\begin{equation}
-5\times log (R/R_{\odot}) -BC -10\times log (T).
\end{equation}
Let us introduce the function $F(CI_0) = BC + 10\times \log (T)$,
the apparent distance modulus $(m-M)_{app} = A + 5\times log
(d/10~pc)$, and rewrite Eq. (2.3) as the light curve model:
\begin{equation}
m = Y - 5\times log (R/R_{\odot}) - F.
\end{equation}
where constant
$$
Y = (m-M)_{app} + M_{{bol}\odot}+10\times log(T_{\odot}).
$$

We now recall that interstellar extinction, $A$, can be determined
from the colour excess $CE$ as $A = R_{\lambda} \times CE$, where
$R_{\lambda}$ is the total-to-selective extinction ratio for the
passband-colour pair considered, whereas $M_{{bol}\odot}$,
$R_{\odot}$, and $T_{\odot}$ are rather precisely known
quantities. The quantity $F(CI_0) = BC + 10\times log (T)$ is a
function of intrinsic colour index $CI_0 = CI - CE$.
\cite[Balona~(1977)]{B77} used a very crude approximation for the
effective temperature and bolometric correction, reducing the
right-hand of the light curve model (2.4) to the linear function
of the observed colour, with the coefficients containing the
colour excess in a latent form.

The key point of our approach is that the values of function $F$
are computed from the already available calibrations of the
bolometric correction $BC(CI_0)$ and the effective temperature
$log T(CI_0)$ ~\cite[(Flower~1996, Bessel, Castelli and Plez~1998,
Alonso, Arribas and Martinez-Roger~1999, Sekiguchi and
Fukugita~2000, Ramirez and Melendez~2005, Biazzo et al.~2007,
Gonzalez Hernandez and Bonifacio~2009)]{F96, BCP98, AAMR99, SF00,
RM05, BFCM07, GHB09}. These calibrations are expressed as
high-order power series in the intrinsic colour:
\begin{equation}\label{Ffun}
F(CI_0) = a_0 + \sum^N_{k=1}a_k \cdot CI_0^k,
\end{equation}
with known $\{a_k\}$ and $N \leq 7$; in some cases the
decomposition also includes the metallicity, $[Fe/H]$, and/or
gravity ($log~g$) terms.

As for the stellar radius, $R$, its current value can be
determined by integrating the star's radial-velocity curve over
time, using $dt = (P/ 2\pi)\cdot d\varphi$:
$$
R(t)-R_0 =-PF \cdot \int^{\varphi}_{\varphi_0}{(V_r(t) -
V_{\gamma})\cdot (P / 2\pi) \cdot d\varphi},
$$
where $R_0$ is the radius value at the phase $\varphi_0$ [we use
mean radius, $<R>=(R_{min}+R_{max})/2$ ]; $V_{\gamma}$ the
systemic radial velocity; $\varphi$ the current phase of the
radial velocity curve, $P$ the star's pulsation period, and $PF$
the projection factor that accounts for the difference between the
pulsation and radial velocities. Given the observables (light
curve and apparent magnitudes, $m$, colour curve and apparent
colour indices, $CI$, and radial velocity curve and $V_r$) and
known quantities for the Sun, we end up with the following
unknowns: distance, $d$, mean radius, $<R>$, and colour excess,
$CE$, which can be found simply using a maximum-likelihood
technique (nonlinear optimization).

For Cepheids with large amplitudes of light and colour curves
($\Delta CI \geq 0.4 mag$), it is also possible to apply a more
general technique by setting the expansion coefficients $\{a_k\}$
in Eq.~(2.5) free and treating them as unknowns. We expanded the
function $F = BC + 10\times log (T)$ in Eq.~(2.4) into a power
series of the intrinsic colour index $CI_0^{st}$ of a well-studied
``standard'' star (e.g., $\alpha$ Per or some other bright star)
with accurately known $T^{st}$:
\begin{equation}
F = BC^{st} + 10\times log (T^{st}) + \sum^N_{k=1}a_k \cdot
(CI-CE-CI_0^{st})^k
\end{equation}

The best fit to the light curve is provided with the optimal
expansion order $N \simeq 5 - 9$. We use this modification to
calculate the physical parameters and reddening $CE$ of the
Cepheids, as well as the calibration $F(CI_0) = BC(CI_0) + 10
\times log (T_{eff} (CI_0))$ for a star of given metallicity
$[Fe/H]$ and $log~g$~cite[(Rastorguev and Dambis~2011)]{RD11}.

\section{Observational data, constants, and best calibrations}

Cepheid photometric data were obtained by L.~Berdnikov; his
extensive multicolor photoelectric and CCD photometry of classical
Cepheids is described in \cite[Berdnikov~(1995)]{Berd95}. Very
accurate radial-velocity measurements of 165 northern Cepheids
were measured with Moscow CORAVEL spectrometer
\cite[(Tokovinin~1987)]{Tok87} during the period 1987-2011
\cite[(Gorynya et al.~1992, 1996, 1998)]{G92, G96, G98}. The total
number of individual measurements is about 11,000 (the latest data
are currently being prepared for publication). The photometric and
spectral data sets are nearly synchronous, to prevent any
systematic errors in the computed radii (up to 30$\%$) and other
parameters owing to evolutionary period changes resulting in phase
shifts between light, colour and radial velocity variations. We
adopt $T_{\odot}$ = 5777 K, $M_{bol \odot} = +4.76^m$
\cite[(Gray~2005)]{G05}. We proceeded from $(V, B-V)$ data and
found as the best solutions for the $V$-band light curve and
$(B-V)$ color curve those computed using the $F[(B-V)_0]$ function
based on two calibrations \cite[(Flower~1996, Bessel, Castelli and
Plez~1998)]{F96, BCP98} of similar slope; the poorer results
obtained using the other cited calibrations can be explained by
the fact that the latter involved an insufficient number of
supergiant stars.

\section{The projection factor}

There is yet no consensus as to which projection factor (PF)
should be used for Cepheid variables ~\cite[(Nardetto et al.~2004,
Groenewegen~2007, Nardetto et al.~2007, Nardetto et
al.~2009)]{Nar04, Gr07, Nar07, Nar09}. Different approaches
(constant or period-dependent $PF$ values) lead to systematic
differences in the inferred Cepheid parameters, first and foremost
in the radii.

Bearing in mind the specific features of CORAVEL measurements,
~\cite[Rastorguev~(2010)]{Ras10} introduced phase-dependent $PF$.
Its value is calculated from flux integration across the stellar
limb and depends on limb darkening coefficient, $\epsilon$ (linear
darkening law $D(\varphi) = 1 - \epsilon + \epsilon \; cos
\varphi, \; \varphi $ is the angle between line-of-sight direction
and the surface element normal vector), and the photospheric
velocity, $dr/dt$. The true line profile will be broadened by any
spectral instrument, and usually we approximate it by Gaussian
curve to measure the radial velocity as the coordinate of the
maximum (see Fig. 1). In this case, the measured radial velocity
will additionally depend on the instrument's spectral linewidth,
$S_0$ (so $PF$ value should be "adjusted" to the spectrograph used
for radial-velocity measurements), and on the photospheric
velocity, $|dr/dt|$.

\begin{figure}
\begin{center}
 \includegraphics[width=3.4in]{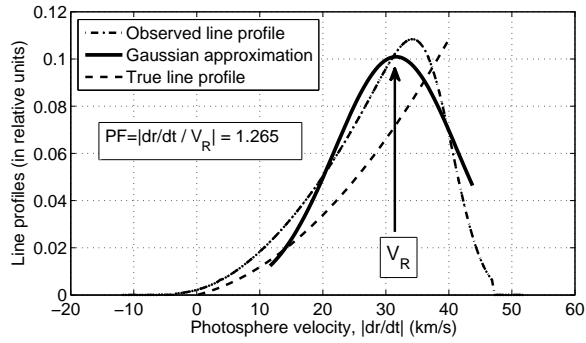}
 \caption{Example: true line profile ($\epsilon = 0.75, dr/dt
= 40 \; km/s$), observed line profile and its Gaussian
approximation (with a spectrograph instrumental line width of $S_0
= 4 \;km/s$). The measured $V_R$ value is indicated by the arrow;
the $PF$ value is shown.}
   \label{fig1}
\end{center}
\end{figure}

\begin{figure}
\begin{center}
 \includegraphics[width=3.4in]{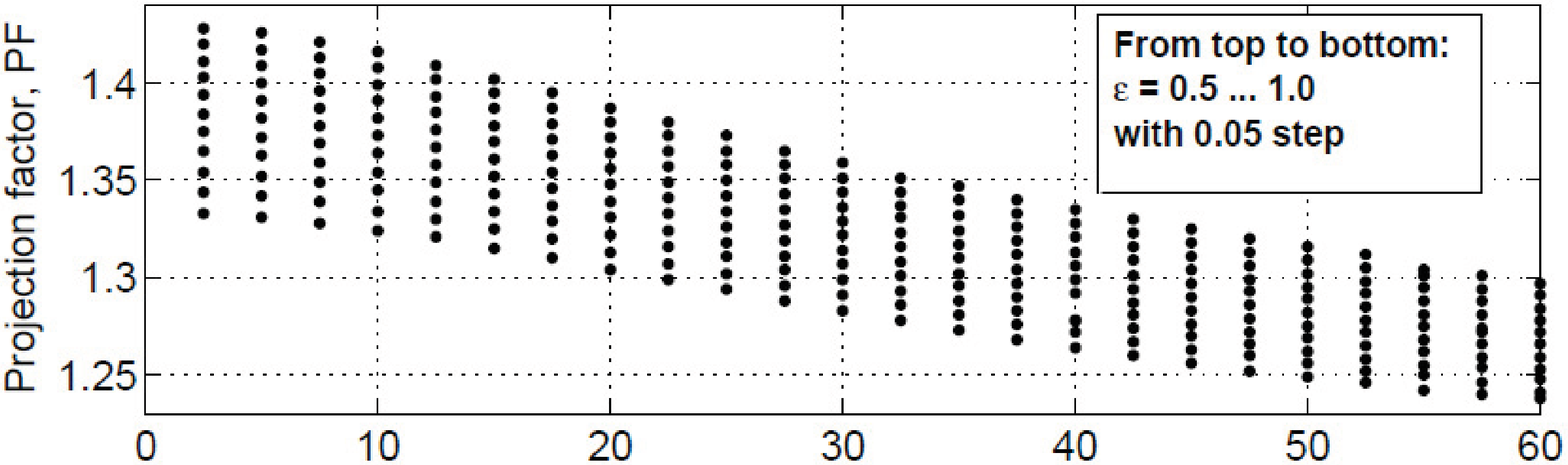}
\caption{Projection factor, $PF$, as a function of $|dr/dt|$ for
an instrumental width of $S_0 = 6 \; km/s$ and for different
values of limb-darkening coefficient $0.5 < \epsilon < 1$.}
   \label{fig2}
\end{center}
\end{figure}

$PF$ values were estimated for $2.5 < |dr/dt| < 60 \; km/s; \; 0 <
\epsilon < 1; \; 4 < S_0 < 8 \;km/s$. We see, that the maximum of
the normal approximation is shifted relative to the tip of the
line profile. This shift depends on $|dr/dt|$ and $S_0$. Fig. 2
shows that the calculated $PF$ values may differ considerably from
the ``standard'' and widely used value $PF=1.31$. Variations of
$PF$ have also been reported by, e.g. Nardetto et al. (2004). We
provide useful analytic approximation for the projection factor,
$PF$, as a three-parameter exponential expression, based on 10,000
numerical experiments ~\cite[(Rastorguev~2010)]{Ras10}:

$PF \approx a_1 \cdot exp[-(dr/dt)^2/(2 \cdot a_2^2)] + a_3$,
where $a_1, a_2, a_3$ are functions of $\epsilon$ and $S_0$:

\(a_1 \approx -0.068 \cdot \epsilon - 0.0078 \cdot S_0 + 0.217\)

\(a_2 \approx +1.69 \cdot \epsilon + 2.477 \cdot S_0 + 9.833\)

\(a_3 \approx -0.121 \cdot \epsilon + 0.009 \cdot S_0 + 1.297\)

Overall, the $rms$ residual for this analytical expression is
approximately 0.003. In practice, the $PF$ value for measured a
given radial velocity should be determined by iteration for known
values of $S_0$. To compare our results on Cepheids with other
calculations, we finally adopted a moderate dependence of $PF$ on
the period advocated by ~\cite[Nardetto et al.~(2007)]{Nar07},
although we repeated all calculations with other variants of $PF$
dependence on the period and pulsation phase to assure the
stability of the calculated colour excess.

\section{Results and discussion}

To test the new method, we used the maximum-likelihood technique
to solve Eq.~(2.3) for the $V$-band light curve and $B-V$ colour
curve for several classical Cepheids residing in young open
clusters: SZ~Tau, CF~Cas, U~Sgr, DL~Cas, and GY~Sge, and found
good agreement for the calculated reddening values with those
determined for the host clusters~\cite[(Rastorguev and
Dambis~2011)]{RD11}. A weak sensitivity of calculated reddening,
$E_{B-V}$, on the adopted $PF$ value (constant or
period/phase-dependent) is explained by the very strong dependence
of the light curve's amplitude on effective temperature, $\sim 10
\times log(T)$, and -- as a consequence -- on the dereddened
colour. Although the internal errors of the reddening $E_{B-V}$
seem to be very small, the values determined using the two best
calibrations ~\cite[(Flower~1996, Bessel, Castelli and
Plez~1998)]{F96, BCP98}, may differ by as much as $0.03 - 0.05^m$,
due to the systematic shift between these two calibrations. Fig.~3
shows the final fit to the $V$-band light curves for some Cepheids
of our sample.

\begin{figure}
\begin{center}
 \includegraphics[width=3.4in]{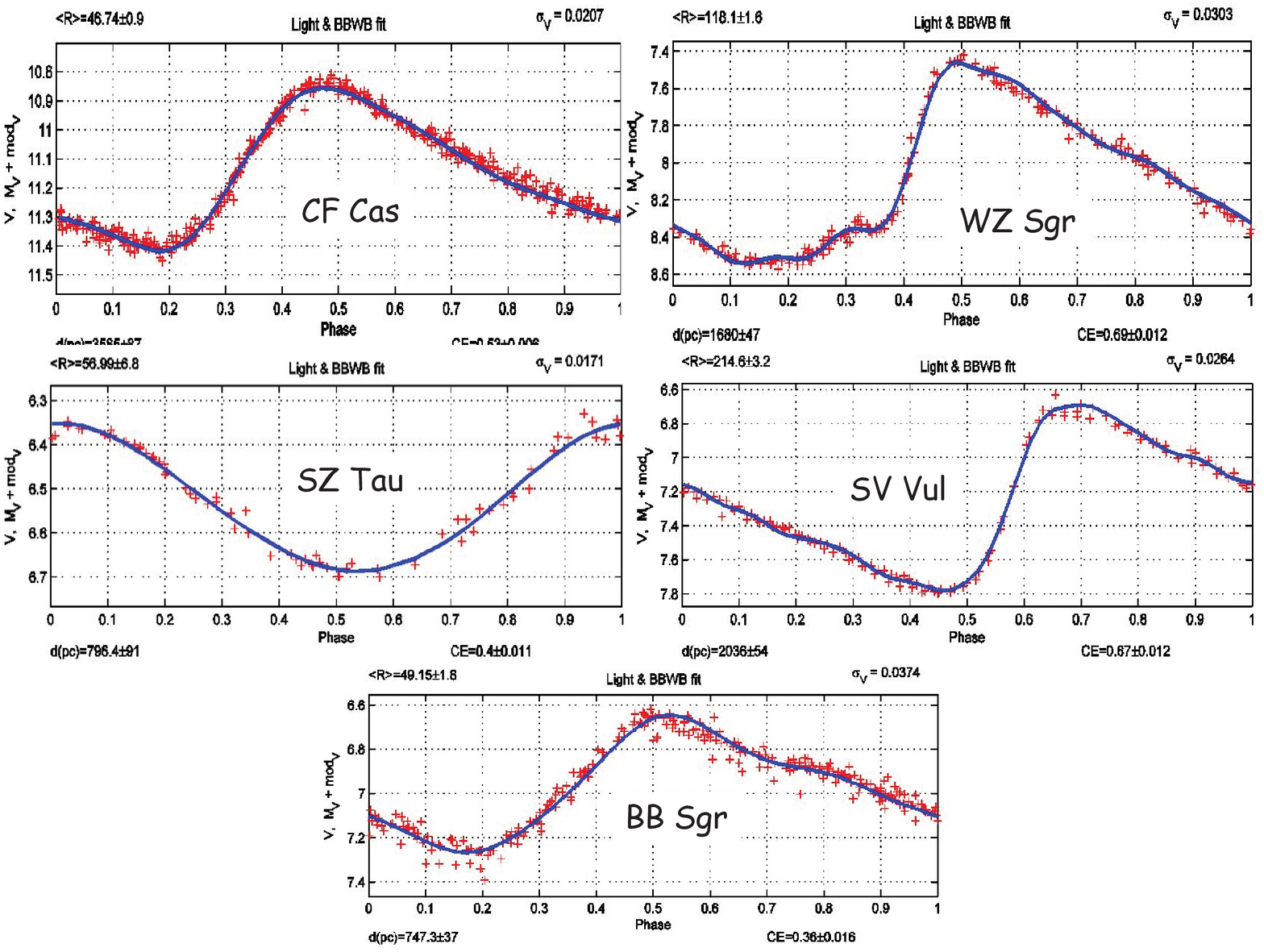}
\caption{Examples of the light-curve modelling. Crosses:
observations; solid line: model. Fit quality: $rms \sim 0.02-0.04
mag$.}
 \label{fig3}
\end{center}
\end{figure}

\begin{figure}
\begin{center}
 \includegraphics[width=5.0in]{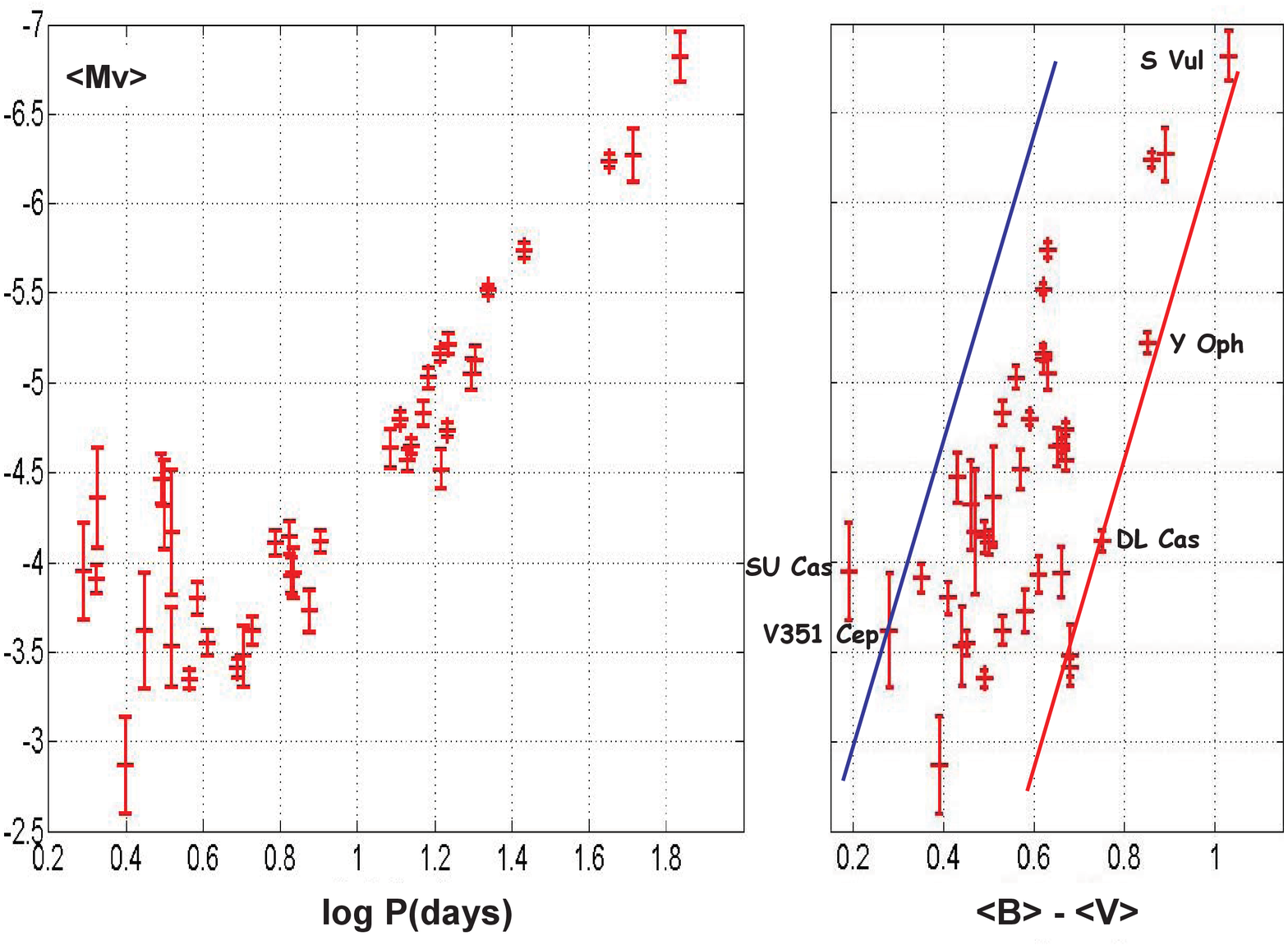}
\caption{PL relation (left) and the instability strip (right) for
approximately 40 Milky Way Cepheids. Intensity mean absolute
magnitudes and colours are shown. Most of short-period Cepheids
seem to be overtone pulsators. Cepheids residing near the edges of
the instability strip and labeled by their names all have very
small amplitudes of the light and colour curves.}
   \label{fig4}
\end{center}
\end{figure}

PL relation and the instability strip for our Cepheid sample are
shown on Fig. 4. Note that the inferred radii and luminosities of
large fraction of the Cepheids with $P < 4^d$ are too large for
fundamental tone pulsations; in most cases this may be indirectly
evidenced by their low colour amplitudes.

To refine the calibration $F(CI_0)$ (2.6), we tried to use
$\alpha$~Per as the ``standard'' star, with $T^{st} \approx
(6240\pm20)~K$, $[Fe/H]\approx -0.28\pm0.06$ \cite[(Lee et
al.~2006)]{Lee06}, $(B-V)^{st}\approx 0.48^m$ and $E_{B-V}\approx
0.09^m$ (WEBDA, for $\alpha$~Per cluster). To take into account
the effect of metallicity on the zero-point $F(CI_0)^{st}$, we
estimated the gradient $dF(CI_0)^{st} / d[Fe/H]\approx +0.24$ from
the calibrations by \cite[Alonso et al.~1999, Sekiguchi \&
Fukugita~2000, Gonzalez Hernandez \& Bonifacio~2009]{AAMR99, SF00,
GHB09}. In some cases (particularly for large-amplitude color
variations) the ``free'' calibration, i.e. Eq.~(2.6), can markedly
improve the model fit to the observed light curve of the Cepheid
variable. Fig. 5 shows the example of calibrations of the $F$
functions derived from nine Cepheids with different $[Fe/H]$ and
$log~g$ values. The temperature scatter at $T_{eff} \sim 6600 -
5100~K$ amounts to $3 - 5\%$.

\begin{figure}
\begin{center}
 \includegraphics[width=4.2in]{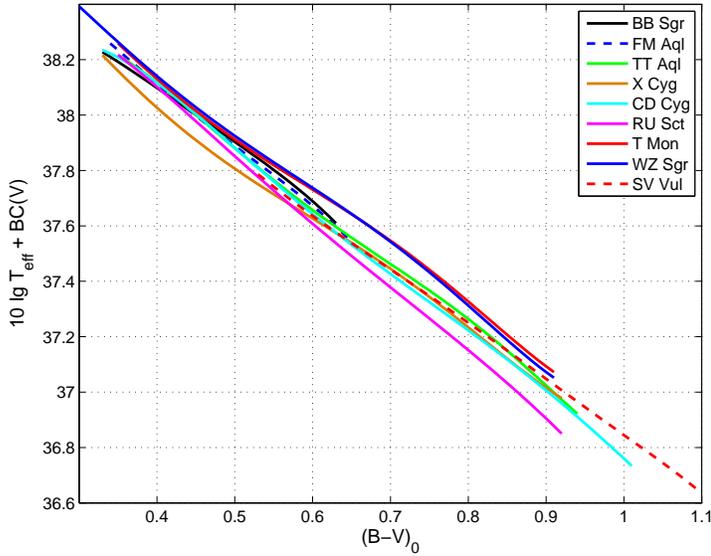}
\caption{Inferred calibrations of the $F((B-V)_0)$ function,
Eq.~(2.6), for nine Cepheids with large amplitudes of their colour
variation.}
   \label{fig5}
\end{center}
\end{figure}

When applied to an extensive sample of Cepheid variables with
homogeneous photometric data and detailed radial-velocity curves,
the new method is expected to lead to a completely independent
scale of Cepheid reddening values and to refine the PL relation.


\begin{thebibliography}{}

\bibitem[Alonso~et~al.~(1999)]{AAMR99} {Alonso, A., Arribas, S., Martinez-Roger, C.} 1999,
\textit{AsApSuppl}, 140, 261

\bibitem[Andrievsky~et~al.~(2002]{Andr02a} {Andrievsky, S. M., Kovtyukh, V. V., Luck, R. E., et
al.} 2002 \textit{A\&A}, 381, 32

\bibitem[Andrievsky~et~al.~(2002]{Andr02b} {Andrievsky, S. M., Kovtyukh, V. V., Luck, R. E., et al.} 2002 \textit{A\&A}, 392,
491

\bibitem[Baade~(1926]{Baa26} {Baade, W.} 1926, \textit{AN}, 228,
359

\bibitem[Balona~(1977)]{B77} {Balona, L. A.} 1976, \textit{MNRAS}, 178, 231

\bibitem[Barnes and Evans~(1976)]{BE76} {Barnes, T. G., Evans, D. S.} 1976,
\textit{MNRAS}, 174, 489

\bibitem[Becker~(1940)]{Beck40} {Becker, W.} 1940, \textit{ZA}, 19,
289

\bibitem[Berdnikov~(1995)]{Berd95} {Berdnikov, L. N.} 1995, In: \textit{Astrophysical
 applications of stellar pulsation. Proc. of IAU Colloq. No.155 held in Cape Town, South
Africa, 6-10 February 1995; eds. Stobie, R. S. and Whitelock,
P.A., Astronomical Society of the Pacific Conference Series}, 83,
349.

\bibitem[Berdnikov~et~al.~(1996)]{Berd96} {Berdnikov, L. N., Vozyakova, O. V., Dambis, A.
K.} 1996, \textit{AstL}, 22, 839

\bibitem[Berdnikov~et~al.~(2000)]{Berd00} {Berdnikov, L. N., Dambis, A. K., Vozyakova, O. V.} 2000, \textit{A\&AS}, 143,
211

\bibitem[Bessell, Castelli and Plez~(1998)]{BCP98} {Bessell, M. S., Castelli, F., Plez,
B.} 1998, \textit{A\&A}, 333, 231

\bibitem[Biazzo et al.~(2007)]{BFCM07} {Biazzo, K., Frasca, A., Catalano, S., et al.} 2007, \textit{AN}, 328,
938

\bibitem[Binney and Merrifield~(1998)]{BM} {Binney, J. and Merrifield,
M.} 1998, \textit{Galactic astronomy, Princeton, NJ : Princeton
University Press}

\bibitem[Dean, Warren and Cousins~(1978)]{DWC78} {Dean, J. F., Warren, P. R., Cousins, A.
W.} 1978, \textit{MNRAS}, 183, 569

\bibitem[Fernie~(1987)]{Fer87} {Fernie, J. D.} 1987, \textit{AJ},
94, 1003

\bibitem[Fernie~(1990)]{Fer90} {Fernie, J. D.} 1990, \textit{ApJ},
354, 295

\bibitem[Fernie~(1994)]{Fer90} {Fernie, J. D.} 1994, \textit{ApJ},
429, 844

\bibitem[Fernie~et~al.~(1995)]{Feretal95} {Fernie, J. D., Evans, N. R., Beattie, B., Seager, S.} 1995, \textit{IBVS},
4148, 1

\bibitem[Fitzpatrick and Massa~(2007)]{FM07} {Fitzpatrick, E. L., Massa, D.}
2007, \textit{ApJ}, 663, 320

\bibitem[Flower~(1996)]{F96} {Flower, Ph. J.} 1996, \textit{ApJ}, 469, 355

\bibitem[Freedman~et~al.~(2001)]{F01} {Freedman, W.~L., Madore, B.~F., Gibson, B.~K.,
et~al.} 2001, \textit{ApJ}, 553, 47

\bibitem[Gonzalez Hernandez and Bonifacio~(2009)]{GHB09} {Gonzalez Hernandez, J. I.,
Bonifacio, P.} 2009, \textit{A\&A}, 497, 497

\bibitem[Gorynya et al.~(1992)]{G92} {Gorynya, N. A.,
Irsmambetova, T. R., Rastorguev, A. S., Samus', N. N.} 1992,
\textit{SvAtL}, 18, 316

\bibitem[Gorynya et al.~(1996)]{G96} {Gorynya, N. A., Samus', N. N.,
Rastorguev, A. S., Sachkov, M. E.} 1996, \textit{AstL}, 22, 175

\bibitem[Gorynya et al.~(1998)]{G98} {Gorynya, N. A., Samus', N. N.,
Sachkov, M. E., Rastorguev, A. S., Glushkova, E. V., Antipin, S.}
1998, \textit{AstL}, 24, 815

\bibitem[Gray~(2005)]{G05} {Gray, C. D. F.} 2005, \textit{The Observation and Analysis of Stellar
Photospheres, Cambridge: Cambridge University Press}

\bibitem[Groenewegen an Oudmaijer~(2000)]{GO00} {Groenewegen, M., Oudmaijer, R.} 2000,
\textit{A\&A} 356, 849

\bibitem[Groenewegen~(2007)]{Gr07} {Groenewegen, M. A. T.} 2007, \textit{A\&A}, 474, 975

\bibitem[Kim~et~al.~(2011)]{KMY11} {Kim, Chulee, Moon, B.-K., Yushchenko, A.
V.} 2011, \textit{JKAS}, 43, 153

\bibitem[Kovtyukh~et~al.~2008]{Kov08} {Kovtyukh, V. V., Soubiran, C., Luck, R. E., et
al.} 2008, \textit{MNRAS}, 389, 1336

\bibitem[Lee et al.~(2006)]{Lee06} {Lee, B.-C., Galazutdinov, G. A., Han, I., et al.} 2006, \textit{PASP}, 118, 636

\bibitem[Madore and Freedman~(1991)]{MF91} {Madore, B. F., Freedman, W.
L.} 1991, \textit{PASP}, 103, 933

\bibitem[Nardetto et al.~(2004)]{Nar04} {Nardetto, N., Fokin, A., Mourard, D., et al.} 2004, \textit{A\&A}, 428, 131

\bibitem[Nardetto et al.~(2007)]{Nar07} {Nardetto, N., Mourard, D., Mathias, Ph., et al.} 2007, \textit{A\&A}, 471, 661

\bibitem[Nardetto et al.~(2009)]{Nar09} {Nardetto, N., Gieren, W., Kervella, P., et al.} 2009, \textit{A\&A}, 502,951

\bibitem[Ramirez and Melendez~(2005)]{RM05} {Ramirez, I., Melendez, J.} 2005,
\textit{ApJ}, 626, 465

\bibitem[Rastorguev~(2010)]{Ras10} {Rastorguev, A. S.}
2010, In: \textit{Variable Stars, the Galactic halo and Galaxy
Formation, Proc. of an international conference held in
Zvenigorod, Russia, 12-16 October 2009; eds. Sterken, Chr., Samus,
N., Szabados, L.}, 225 (see also Rastorguev, A. S. ``Variable
stars, distance scale, globular clusters" arXiv:1001.1648v2)

\bibitem[Rastorguev and Dambis~(2011)]{RD11} {Rastorguev, A.~S., Dambis, A.~K.} 2011,
\textit{AstrBull}, 66, 47

\bibitem[Sandage~et~al.(2006)]{Sand06} {Sandage, A., Tammann, G. A., Saha, A., et
al.} 2006, \textit{ApJ}, 653, 843

\bibitem[Sekiguchi and Fukugita~(2000)]{SF00} {Sekiguchi, M., Fukugita, M.} 2000,
\textit{AJ}, 120, 1072

\bibitem[Tokovinin~(1987)]{Tok87} {Tokovinin, A. A.} 1987,
\textit{SvA}, 31, 98

\bibitem[van Leewen~2007]{vL07} {van Leeuwen, F., Feast, M. W., Whitelock, P. A., et al.} 2007, \textit{MNRAS}, 379, 723

\bibitem[Wesselink~(1946)]{Wes46} {Wesselink, A.J.} 1946, \textit{BAN}, 10, 91

\end{thebibliography}
\end{document}